\documentclass[manuscript]{acmart}

\AtBeginDocument{%
  \providecommand\BibTeX{{%
    \normalfont B\kern-0.5em{\scshape i\kern-0.25em b}\kern-0.8em\TeX}}}

\setcopyright{acmcopyright}
\copyrightyear{2018}
\acmYear{2018}
\acmDOI{XXXXXXX.XXXXXXX}

\acmJournal{JACM}
\acmVolume{37}
\acmNumber{4}
\acmArticle{111}
\acmMonth{8}

\acmPrice{15.00}
\acmISBN{978-1-4503-XXXX-X/18/06}

\usepackage{subcaption}

\begin{document}

\newcommand{\subsubsubsection}[1]{\textbf{#1.}}

\title[Understanding Motivations, Mental Models, and Concerns of Users Flagging Social Media Content]{Cleaning Up the Streets:  Understanding Motivations, Mental Models, and Concerns of Users Flagging Social Media Content}

\author{Alice Qian Zhang}
\email{zhan6698@umn.edu}
\affiliation{%
    \authornote{This work was written while the author was a student at the University of Minnesota.}
  \institution{Carnegie Mellon University}
  \city{Pittsburgh, PA}
  \country{USA}
  }

\author{Kaitlin Montague}
\email{kaitlin.montague@rutgers.edu}
\affiliation{%
  \institution{Rutgers University}
  \city{New Brunswick, NJ}
  \country{USA}
  }  

\author{Shagun Jhaver}
\email{shagun.jhaver@rutgers.edu}
\affiliation{%
  \institution{Rutgers University}
  \city{New Brunswick, NJ}
  \country{USA}
  }

\renewcommand{\shortauthors}{Zhang, et al.}

\begin{abstract}
Social media platforms offer flagging, a technical feature that empowers users to report inappropriate posts or bad actors to reduce online harm. The deceptively simple flagging interfaces on nearly all major social media platforms disguise complex underlying interactions among users, algorithms, and moderators. Through interviewing 25 social media users with prior flagging experience, most of whom belong to marginalized groups, we examine end-users' understanding of flagging procedures, explore the factors that motivate them to flag, and surface their cognitive and privacy concerns. We found that a lack of procedural transparency in flagging mechanisms creates gaps in users' mental models, yet they strongly believe that platforms must provide flagging options. Our findings highlight how flags raise critical questions about distributing labor and responsibility between platforms and users for addressing online harm. We recommend innovations in the flagging design space that enhance user comprehension, ensure privacy, and reduce cognitive burdens. 
\end{abstract}

\ccsdesc[500]{Human-centered computing~Empirical studies in collaborative and social computing}

\keywords{content moderation, flags, transparency}

\maketitle
\section{Introduction}
End-users have several options to take action when encountering inappropriate content on social media platforms, from regular interactions built into a platform (e.g., unfollowing an account) to content moderation-specific interactions. Of these options, it is only through \textbf{flagging} that end-users can directly request platform administrators to take site-wide actions against the content. 
Platforms maintain content regulation systems (comprising a coordinated deployment of automated tools and human reviewers~\cite{jhaver2019automated,Gorwa2020}) that regularly review flagged items and, when warranted, trigger sanctions, such as removing flagged posts or banning flagged accounts~\cite{goldman2021content,grimmelmann2015virtues}.
Thus, the \textit{flag} serves as a powerful and empowering tool for user-driven moderation.  

Though at first glance, flag implementations may signal platforms' commitment to enacting democratic governance and fostering user communities, in practice, flags are often designed without adequately considering the needs of users who rely on them. As Crawford and Gillespie describe, flagging is a \textit{``complex interplay between users and platforms, humans and algorithms, and the social norms and regulatory structures of social media''} ~\cite{crawford2016}. However, while their work identified the conceptual significance of flags as critical socio-technical mechanisms, there is little empirical research examining how users---the primary stakeholders in the flagging processes---experience and perceive these systems. This lack of focus on user perspectives is particularly important given that flagging processes across platforms significantly vary regarding the input required of flag submitters and post-flagging review and notification procedures. 

\begin{figure}
  \centering
  \begin{subfigure}{0.3\textwidth}
    \centering
    \includegraphics[width=0.5\linewidth]{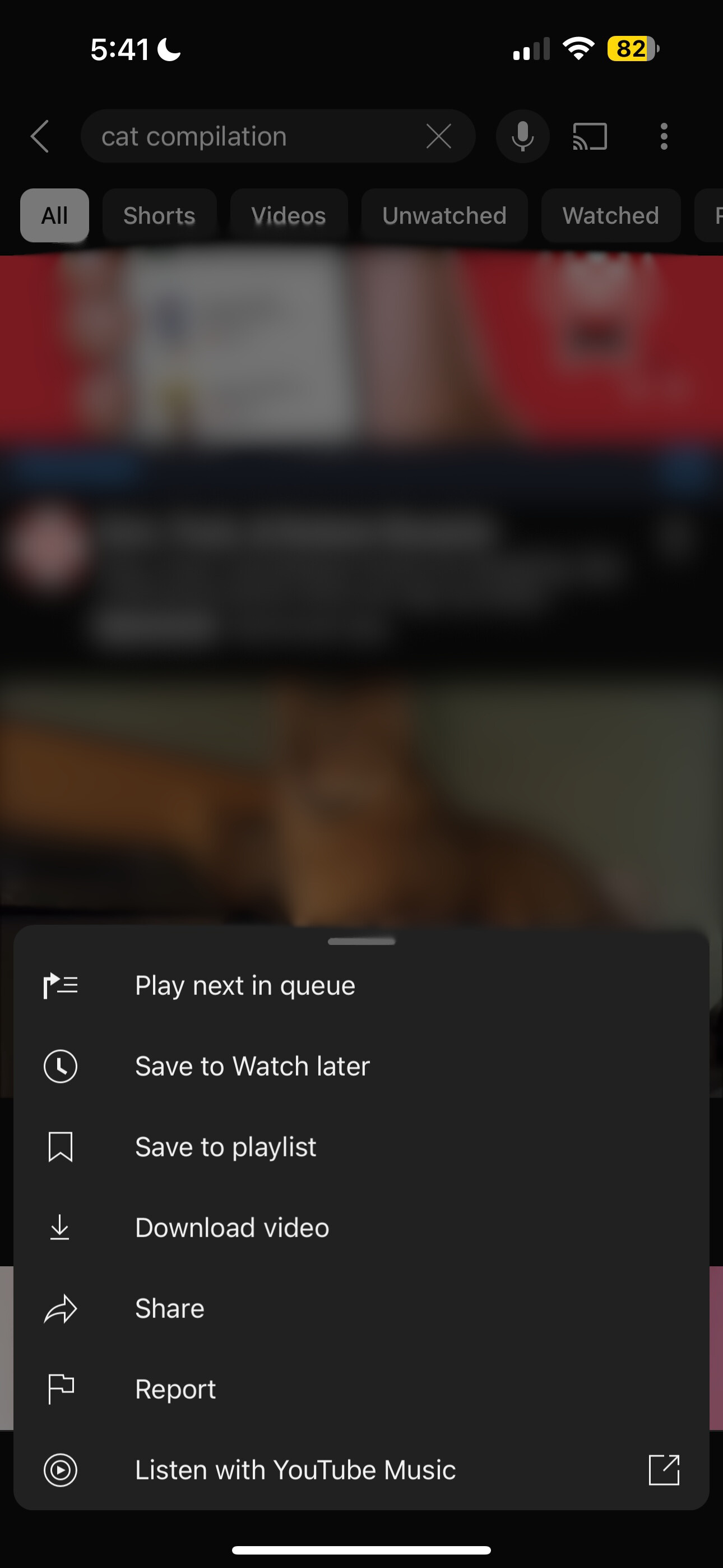}
    \caption{The report button for flagging content on Youtube.}
    \label{fig:sub1}
  \end{subfigure}\hfill
  \begin{subfigure}{0.3\textwidth}
    \centering
    \includegraphics[width=0.5\linewidth]{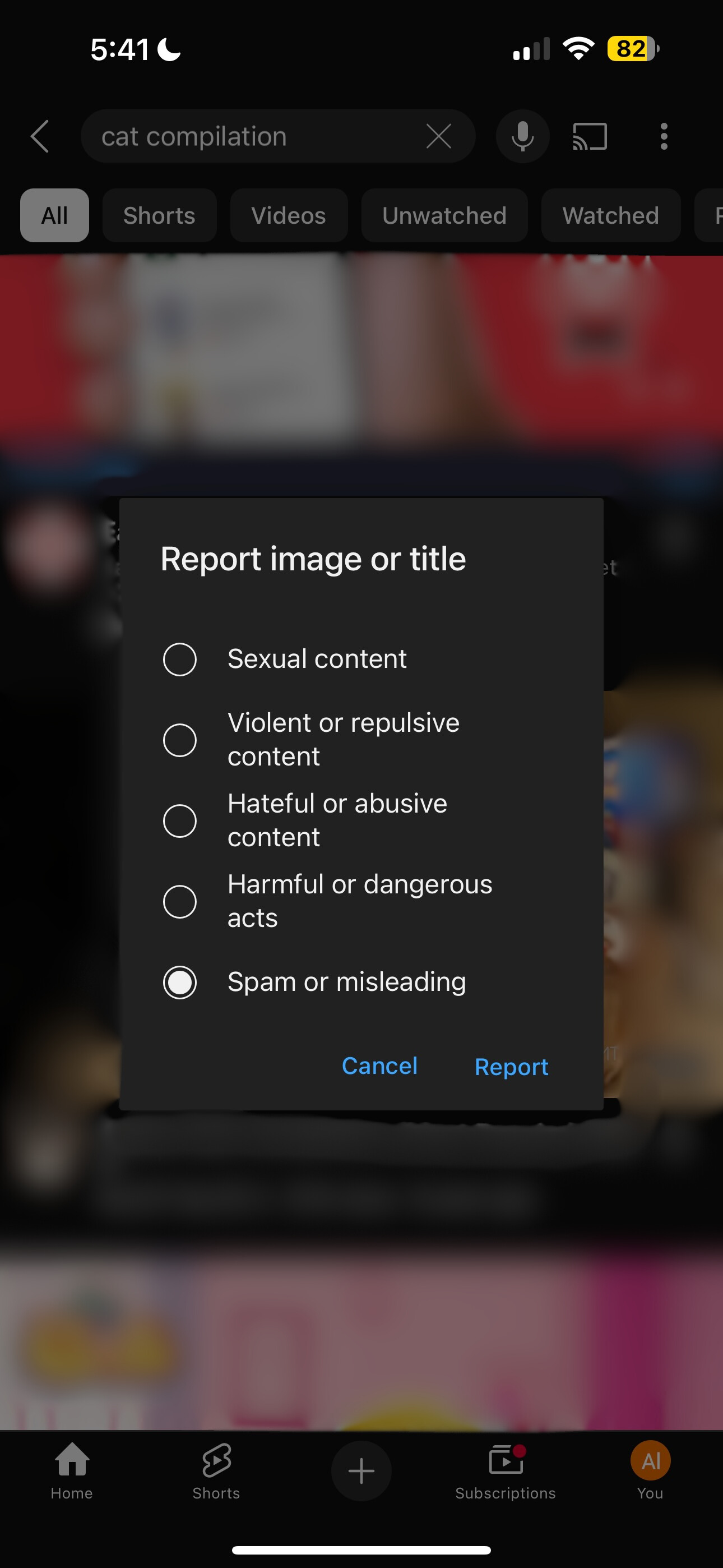}
    \caption{Categories for flagging content on Youtube.}
    \label{fig:sub2}
  \end{subfigure}\hfill
  \begin{subfigure}{0.3\textwidth}
    \centering
    \includegraphics[width=0.5\linewidth]{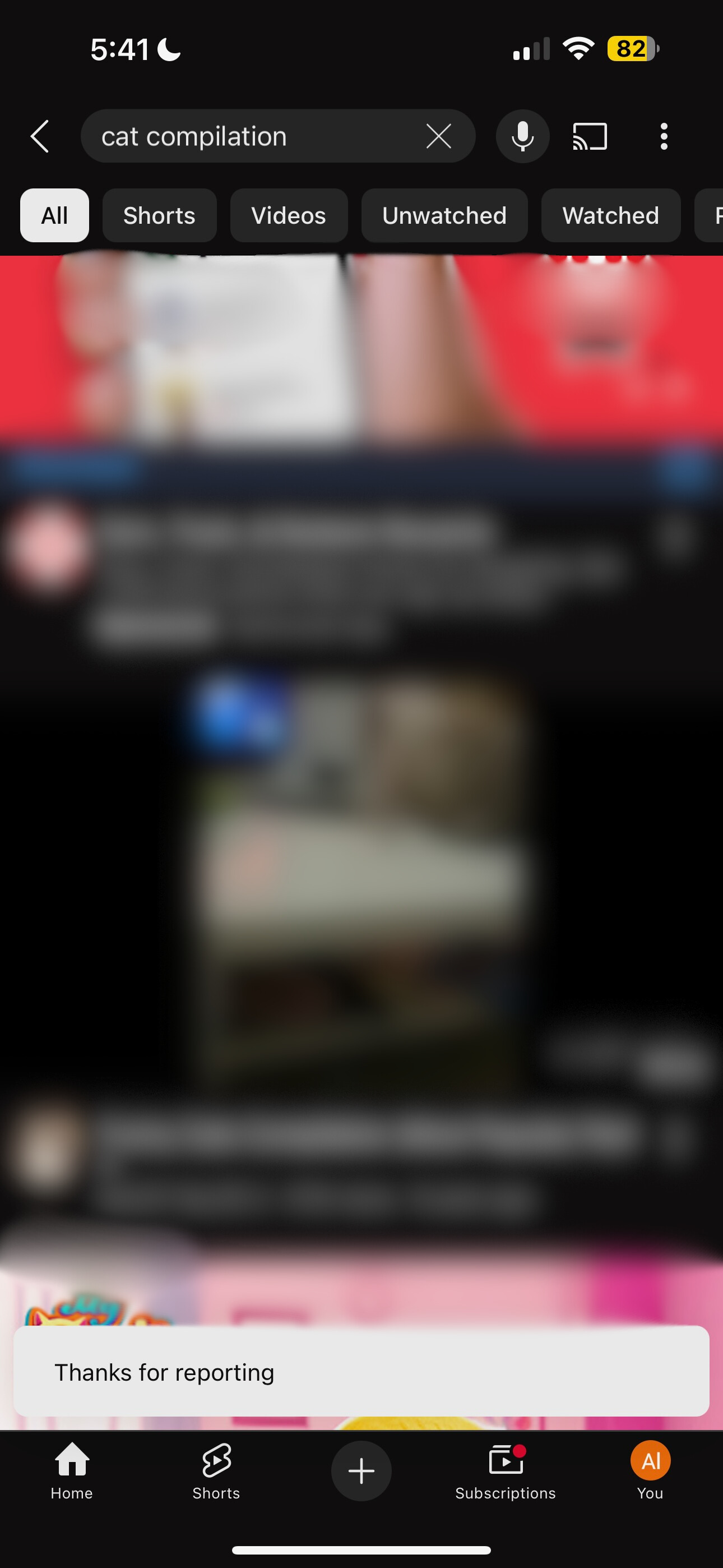} 
    \caption{Thank you pop-up after users flag content on Youtube.}
    \label{fig:sub3}
  \end{subfigure}
  \caption{YouTube's flagging interface pages.}
  \label{fig:youtube_process}
\end{figure}
\begin{figure}
\centering
\includegraphics[width=0.8\textwidth]{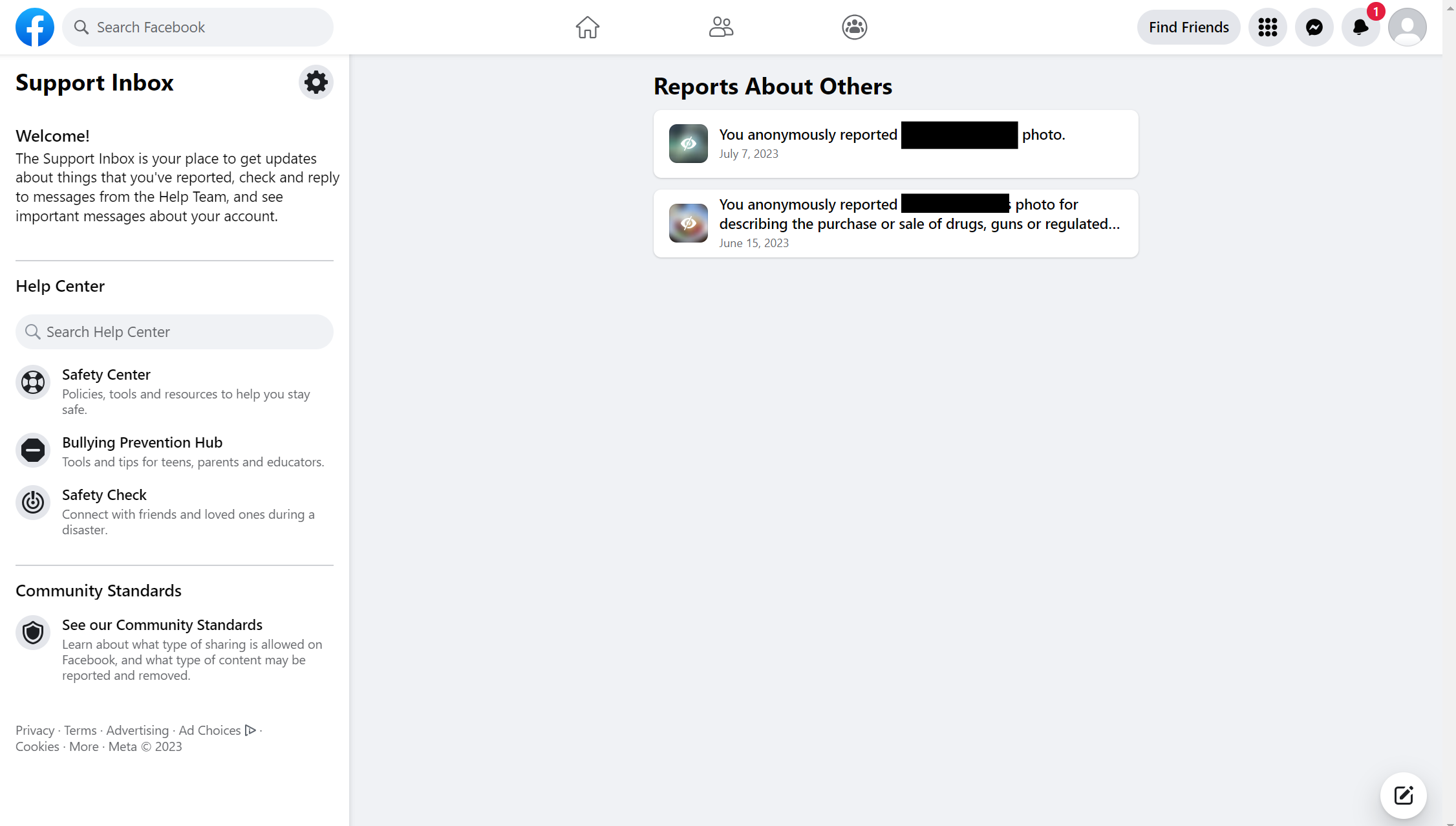}
\caption{Facebook's support inbox dashboard showing posts a user has flagged.}
\label{fig:facebook_process}
\end{figure}

For instance, when a user flags a video on YouTube, they see a quick pop-up that thanks users for flagging content with no other contact from the platform about the outcome of their flag (Fig. \ref{fig:youtube_process}). In contrast, when users flag content on Facebook, the platform typically tells them they will be notified when a decision is made. It provides them with additional safety options such as blocking, muting, and hiding the flagged content. Facebook also offers users a centralized interface where they may revisit past flags to check their review status (Fig. \ref{fig:facebook_process}). These different implementations suggest that current flagging systems are frequently designed in an ad hoc fashion, and it remains unclear if they adequately understand user needs despite users being central to their success.

In this study, we analyze user experiences with flagging mechanisms on major social media platforms and ask three key research questions.
First, we note that flagging interfaces provide varying degrees of information regarding how the user input during flagging will be processed, how long it will take, who will review the flag, and more.
It is unclear how users interpret these information cues and what shapes their understanding. Therefore, we ask:

\begin{quote}
\textbf{RQ1:} \textit{What are users’ mental models of how social media platforms implement flagging mechanisms, and how do they develop these models?}
\label{quote:RQ1}
\end{quote}

Next, we examine why users choose to flag content on social media platforms. While flagging mechanisms have been studied in other contexts, such as gaming platforms, where users may flag for reasons unrelated to platform-defined toxicity (e.g., flagging teammates for poor game performance \cite{kou2021flag}), there is limited understanding of why users flag harmful content on social media. Social media platforms are unique in their role as central hubs for public discourse, making it critical to understand the factors driving users to flag or choose to not flag harmful content. Investigating these motivations is essential for ensuring that flagging mechanisms are designed to meet users' needs and align with platform goals. Thus, we ask:

\begin{quote}
\textbf{RQ2:} \textit{What motivates users to engage (or not engage) in flagging on social media platforms?}
\label{quote:RQ2}
\end{quote}

Submitting a flag is an act of anonymous and voluntary labor that requires users to navigate a multi-step interface, where they must categorize their concerns under the platform’s pre-defined labels of inappropriate conduct. This process not only demands cognitive effort to evaluate and classify content but also raises questions about users' privacy and potential risks. To better understand these challenges, we investigate the cognitive and privacy-related implications of flagging by asking:
\begin{quote}
\textbf{RQ3:} \textit{What are users’ cognitive and security concerns about engaging in flagging on social media platforms?}
\label{quote:RQ3}
\end{quote}

To answer these questions, we conducted semi-structured interviews with 25 social media users who recently flagged content. We first asked participants about their experiences of flagging inappropriate content online. Next, we asked questions that probed their understanding of how flagging works. Finally, we examined participants' perspectives on possible changes to flagging mechanisms, especially regarding their transparency and feedback. 

Applying interpretive qualitative analysis~\cite{merriam2002} to our interview data, we found that flagging a post involves three temporally distinct stages from the flagger's perspective: (1) before, (2) during, and (3) after flagging. However, within each of these stages, the lack of transparency in flagging creates substantial gaps in users' mental models about how platforms evaluate flagged content. Given this lack of transparency, users develop strategies to leverage their interpretation of how flags work, using flags alongside other site mechanisms, such as blocking and private messaging, to address online harms. A belief in generalized reciprocity motivates users to flag inappropriate content---they expect a collective flagging effort would help keep online spaces clean. However, flagging is not without its costs. We surface the reservations users have about the cognitive burdens of flagging and examine their privacy concerns when discussing potential innovations to flagging mechanisms.

Our main contributions are:
\begin{enumerate}
    \item \textit{Empirical data on user experiences and perspectives about flagging on social media platforms.} We provide empirical data contributing to Crawford and Gillespie's theoretical conceptualization of flagging ~\cite{crawford2016}. Our findings highlight the importance of flags on social media as a vital way for users to voice concerns within an opaque content moderation system. We also explore views of flags as a right and obligation, surfacing tensions between user motivations to flag and platform motives for offering flags. We provide insights about the labor of flagging. 
    
    \item \textit{Conceptualization of the three stages of flagging (before, during, and after).} We provide a conceptual understanding of flagging from end users' perspectives across these three temporal stages. We also highlight the key user concerns within each stage.
    
    \item \textit{Design opportunities with recommendations.} We provide empirically informed design recommendations and identify opportunities for innovation in building flagging systems. We recommend incorporating seams~\cite{eslami2016first} into flagging interfaces to help users understand their operations and ensure transparency throughout the flagging process. We recommend that innovations to flagging mechanisms should address user needs while respecting privacy and cognitive burdens. 
\end{enumerate}
 
\section{Related Work}
\subsection{Content Moderation Affordances on Social Media}
Most contemporary platforms hosting user-generated content offer a mechanism for users to provide feedback on the content they see that violates the platform's guidelines. This mechanism is usually called ``flag'' or ``report.'' Platforms use flags to identify potentially inappropriate content, which human moderators or bots subsequently review ~\cite{kou2021flag}. If this review determines that the content indeed violates the platform's guidelines, the flagged user may face direct consequences, such as content removal and account suspension ~\cite{jhaver2019survey, jhaver2021evaluating}. 

We contextualize flags as part of a collection of mechanisms that let users express their voice in content moderation systems. These mechanisms commonly involve expressing dissatisfaction about norm-violating content (e.g., hate speech, violence, spam) ~\cite{kumar2021designing, banko2020unified, scheuerman2021framework} or behaviors (e.g., harassment, bullying, scamming) ~\cite{shaw2018beyond}. 
Flags are often accompanied by a variety of \textit{personal moderation tools}, which have been defined as ``tools that let users configure or customize some aspects of their moderation preferences on social media'' ~\cite{jhaver2023personalizing}. 
These personal moderation tools include \textit{account-based tools} that let users mute or block specific accounts \cite{geiger2016,matias2015reporting,jhaver2018blocklists} and \textit{content-based tools} that let users hide comments containing configured keywords ~\cite{jhaver2022filterbuddy} or set up content sensitivity controls ~\cite{jhaver2023users}. We focus on flagging, one of the most conspicuous content moderation mechanisms available to general end-users. While personal moderation tools change only the configuring user's news feed, flags enable site-wide content regulation, provide higher granularity in letting users express their dissatisfaction, and prompt users to justify their actions. \textit{Appeal mechanisms} ~\cite{vaccaro2020at, vaccaro2021contestability}, which let moderated users report their dissatisfaction with content moderation decisions and request that they be reversed, are often used to contest platform moderation decisions about flagged content or other content removed by administrators. Users may also utilize the above moderation mechanisms in combination with regular platform interactions that demonstrate positive or negative feedback (e.g., like and dislike buttons)~\cite{khan2017social} or other social signals (e.g., unfollowing accounts) ~\cite{sasahara2021social}. 

Users with special roles (e.g., community moderators) usually have access to a wider range of moderation tools ~\cite{Hatmaker_2021, stockinger2023navigating}. For example, Reddit moderators may independently program Automod, a bot, to automatically identify and take action against content that infringes upon their community's guidelines~\cite{wright2022automated,jhaver2019automated}.
Similarly, Twitch streamers and YouTube creators rely on third-party applications for harmful content detection~\cite{cai2019categorizing,jhaver2022filterbuddy}. We do not focus on such specialized tools because this study is examining regular end-users who lack access to them. However, we draw from literature on accountability and division of labor involved in the deployment of such tools ~\cite{jhaver2019automated,anaobi2023admins,chandrasekharan2019crossmod} to examine how end users perceive the use of automation and human moderators to review flagged content.

\subsection{User Perspectives on Content Moderation}
Prior work has asserted the importance of examining user perspectives of content moderation to inform platform practices and policymaking. Scholars argue that these internet governance regimes are defining users' roles and shaping our political and public discourse, so it is imperative to incorporate users' perspectives in their design ~\cite{riedl2021responsible,langvardt2017regulating}. Our research on users' perceptions of flagging mechanisms highlights existing user needs to inform changes to how platforms design and administer these systems. 

The discussion of user perspectives includes questions about who owns the responsibility for platform governance. Gillespie advocated for a collective civic responsibility to govern platforms ~\cite{gillespie2018custodians}. However, it is not always clear how we may distinguish specific stakeholder roles when applying this form of collective governance. For instance, prior research shows that users attribute the responsibility for interventions and detection of inappropriate content to platforms \cite{riedl2021responsible} or human moderators  ~\cite{naab2018flagging} rather than themselves. However, ~\citet{jhaver2023users} showed that users prefer personally configurable moderation tools to regulate hate speech, violent content, and sexually explicit content instead of platform-enacted bans of those categories.
This suggests that users desire to shape their feeds but expect platforms to intervene when necessary.
We build on this work by examining how end-users perceive their responsibilities to flag fitting in with how they assume the obligations to regulate inappropriate content are divided among all stakeholders. 

Prior research has highlighted substantial concerns about the burdens of physical and emotional labor that moderation places on regular end-users ~\cite{jhaver2023personalizing}, volunteer moderators~\cite{li2022all, jhaver2019automated,wohn2019volunteer, dosono2019mod, ruckenstein2020re}, and commercial moderators ~\cite{ roberts2019behind, steiger2021psy} and the tensions of how users' volunteer efforts factor into platforms' monetary gains ~\cite{wohn2019volunteer,li2022measuring}. While the emotional burden of moderation on volunteer and hired human moderators has been examined in detail~\cite{dosono2019mod,roberts2019behind}, we find that scant prior literature has observed the additional burden on regular users of reporting content and thus aim to explore this aspect in our study.

As an extension of discussions regarding the responsibilities to flag harmful content and the burden it places on users, it is known that people of marginalized identities are more likely to experience harm on online platforms~\cite{duggan_online_2017}. For example, Black and female-identifying users were shown to disproportionately experience an array of harms from unwanted behaviors on platforms (e.g., hate speech, doxxing) ~\cite{musgrave2022experiences}. To address these challenges, HCI and CSCW researchers have prioritized the content moderation needs and perspectives of vulnerable groups~\cite{haimson2021disproportionate,thach2024visible, liang2021embracing}.
For example, Blackwell et al. examined HeartMob users' experiences on a platform built for those most affected by severe online abuse ~\cite{blackwell2017classification}. They argue that platforms' design and moderation must integrate vulnerable users' unique needs to ``fully'' address online harassment. This is also the rationale for our emphasis on examining the diverse experiences of users from marginalized identities with flagging mechanisms. We explore participants' concerns about flagging with a particular focus on fears of retaliation, concerns about privacy, and misuse of flagging mechanisms.

When considering the design of moderation systems, it is also important to consider end-users' preferences for control and transparency. Currently, most social media platforms do not offer adequate visibility into the rules and the decision-making processes behind their moderation decisions ~\cite{suzor2019we}.
Sometimes, transparency challenges are a result of issues with existing moderation interfaces that, for example, do not allow for transparency about what decisions are made by automated tools versus human moderators ~\cite{ozanne2022shall, juneja2020through}. Some researchers have examined how different configurations of moderation tool design shape the control that users perceive over their social media feeds~\cite{jhaver2022filterbuddy,jhaver2023personalizing}. Additionally, prior research shows that when moderation systems indicate the reasoning behind content regulation decisions, user attitude toward those decisions may improve, resulting in higher-quality user posts ~\cite{jhaver2019explanations, jhaver2019survey, kim2009} or generally increased trust  ~\cite{brunk2019effect}. Flagging systems similarly lack transparency as platforms are often unclear about when they act on flagged content and what actions they perform. Such practices result in user perspectives being overlooked and unheard. For this reason, we examine users' transparency needs and offer empirical insights about how they seek individual agency in flagging processes.

\subsection{Flagging Mechanisms on Social Media Platforms}
While it is unclear when flags were first implemented on social media platforms, the notion of raising a concern to a community leader or administrator has long been used to express dissatisfaction civilly~\cite{dibbell1994rape}. Flags are a crucial means for users to express such dissatisfaction in online environments. In this section, we provide an overview of related research on flagging as well as a description of flagging mechanisms on select social media platforms to better situate the context of our study.

Previous research examined flagging from end-users' perspectives as creators of sanctioned content, especially in the context of how users develop ``folk theories'' \cite{devito2017algorithms} about moderation decisions when platforms fail to provide adequate or any explanations for those decisions~\cite{thach2024visible}. The opaqueness of flagging has led to speculations about platforms making controversial decisions about flagged content without the flagged user's knowledge, such as with shadowbanning (where platforms hide a user's content from others' feeds without removing it entirely) ~\cite{myers2018censored}. This has raised concerns about the use of flagging to silence \cite{are2023flagging} or otherwise censor users~\cite{peterson2013user}. Our study expands on this work by specifically examining the perspectives of users when they are flagging others' content rather than being flagged themselves.

The definition of how ``flaggable'' a post is, or its ``flaggability,'' for users may differ from platforms' definitions of what is inappropriate. For example, users on gaming platforms utilized flags to report what they defined as toxic content but also flagged teammates to whom they attributed game losses~\cite{kou2021flag}. 
\citet{Zhao2023Chinese} reported a case study of mass reporting on Webo where fans of Chinese celebrities publicly coordinated to report content critical of those celebrities in an effort to take it down.
What's more, what is interpreted as harmful enough to flag may differ from user to user~\cite{jhaver2018view,luo2024participatory}, and platforms have differing guidelines on what is deemed inappropriate and thus flaggable~\cite{pater2016}. These findings highlight a gap in our understanding of how users navigate conflicts between their values and platform guidelines for flagging. It also remains unclear how such appropriations of flags occur on social media platforms. Thus, we examine the decision-making process behind how users choose to flag inappropriate content online. Drawing from prior literature, we refer to inappropriate content as a multitude of behaviors~\cite{scheuerman2021framework} that include but are not limited to trolling ~\cite{gerlitz2013like,phillips2015we}, invasion of privacy ~\cite{sambasivan2019they, chen2023research, boyd2008facebook, aimeur2011ultimate}, public shaming ~\cite{vaccaro2021contestability,ronson2015, goldman2015trending}, and interpersonal harm~\cite{riedl2021responsible}.

\citet{wang2023reporting} examined users' privacy concerns when reporting messages or accounts perpetrating harassment on end-to-end encrypted messaging platforms. They found that users make nuanced decisions about whether and how much to report based on their perceived trade-offs between privacy risks and protections. Further, these decisions are influenced by their trust in messaging platforms and community moderators. We build on this work to examine social media users' privacy concerns when they consider reporting inappropriate content. \citet{riedl2024reporting} reported a case study of Taylor Swift fans reporting nonconsensual pornographic deepfakes of the pop superstar on X. They noted that such reporting is motivated by fans' strong sense of responsibility to protect Taylor Swift online. We add to this research by examining how users perceive their versus platforms' flagging obligations more broadly.


At the moment, major social media platforms allow users to flag content, but post-flagging actions vary widely--from no updates on removal decisions (YouTube, Fig. \ref{fig:youtube_process}) to automatically blocking accounts for users who flag them (X, formerly Twitter) and offering the option to unfollow or block accounts (Instagram and TikTok) to letting users track flagged content through dashboards (Facebook, Fig. \ref{fig:facebook_process}). Despite the prevalence of flagging, research has primarily focused on other moderation strategies~\cite{katsaros2022reconsidering,chang2019trajectories,jhaver2023personalizing}. Our study fills this gap by analyzing the user experience of flagging mechanisms across platforms, offering empirical insights to design flagging systems that better support users in addressing inappropriate content.

\section{Methods}

\subsection{Participant Recruitment}
Our university's IRB\footnote{We will specify the University name after the anonymous peer review is completed.} approved this study on Jan 31, 2023. 
We recruited participants by posting on Twitter, NextDoor, and Craigslist. In selecting these platforms to place our recruitment calls on, we aimed to reach out to a diverse set of potential participants. We asked candidates to submit an online form that included questions about whether they use social media daily, which social media sites they use, whether they have encountered toxic content on social media, whether they flagged a post in the past week, and their demographic information. We included an option to attach a screenshot for a post the users recently flagged. We also included an open-ended question: ``What is your perspective on how social media platforms can improve flagging mechanisms?''

We reviewed submissions and screened for responses with legitimate and thoughtful answers. 
In selecting interview candidates, we drew inspiration from prior literature that advocates for researchers and practitioners to give voice to the perspectives of users who are more likely to be vulnerable to online harm ~\cite{duggan_online_2017, musgrave2022experiences, blackwell2017classification} as well as literature showing such users are more likely to be exposed to content that is harmful~\cite{duggan_online_2017,jhaver2018blocklists}. Additionally, we note prior examples of research on user perspectives on content moderation that also oversampled users with marginalized identities ~\cite{jhaver2023personalizing}.
Therefore, we prioritized offering interviews to candidates who self-identified as belonging to marginalized groups, such as Black and LGBTQ+ users. Table \ref{table:participants} shows the demographic details of our participants.

All participants indicated that they had flagged at least one social media post within the past month of their interview date. We used this experience of flagging to ground our interview discussions and let participants provide more specific feedback to our queries about their understanding of the purpose and operations of flags, their motivations for flagging, and their concerns with flagging. 

\begin{table}
\footnotesize
 \hspace*{-0.8cm} 
\begin{tabular}{|l|l|l|p{1.3cm}|l|p{1.1cm}|l|p{2.8cm}|}
\hline
\textbf{\#} & \textbf{Age} & \textbf{Gender} & \textbf{Race} & \textbf{Occupation}              & \textbf{Country}      & \textbf{Social Media Platforms Used} \\ \hline 
P1          & 35           & Non-Binary   & White   & Digital Content Producer         & USA                   & Facebook, Instagram, Twitter, Tiktok, Snapchat          \\ \hline        
P2          & 32           & Female      & Black    & Fashion Designer                 & USA                   & Facebook, Instagram, Twitter            \\ \hline             
P3          & 53           & Female     & White     & Professor                        & USA                   & Facebook, Twitter           \\ \hline              
P4          & 27           & Female     & White         & PhD Researcher                   & France                & Twitter, Whatsapp            \\ \hline           
P5          & 33           & Non-Binary & Native American         & Editor                           & USA                   & Facebook, Instagram, Twitter, Tumblr            \\ \hline            
P6          & 22           & Female     & Black         & Waitress                         & Canada                & Facebook           \\ \hline          
P7          & 24           & Male       & Black        & Artist                           & USA                   & Facebook, Instagram            \\ \hline              
P8          & 27           & Male       & Black         & Carpenter                        & USA                   & Facebook, Twitter, Snapchat            \\ \hline             
P9          & 22           & Male       & Black         & Fashion Designer                 & USA                   & Facebook, Twitter            \\ \hline              
P10         & 30           & Male       & Black        & Electrician                      & USA                   & Facebook, Instagram            \\ \hline              
P11         & 26           & Male       & Black        & Plumber                          & USA                   & Instagram, TikTok, Snapchat            \\ \hline              
P12         & 20           & Female     & Asian         & Student                          & Canada                & Facebook, Instagram, Youtube, Reddit             \\ \hline                               
P13         & 24           & Female     & Black         & Student                          & USA                   & Instagram, Twitter            \\ \hline               
P14         & 27           & Male       & Black         & Uber Driver                      & The Netherlands       & Facebook, Instagram, Twitter            \\ \hline  
P15         & 26           & Male       & Black         & Sales Representative             & Belgium               & Facebook, Instagram, Pinterest            \\ \hline 
P16         & 35           & Male       & Black         & Engineer                         & UK                    & Facebook, Instagram, Whatsapp           \\ \hline       
P17         & 27           & Male       & Black         & Accountant                       & UK                    & Facebook, Twitter, Reddit            \\ \hline        
P18         & 25           & Male       & Black         & Sales Agent                      & USA                   & Facebook, Twitter            \\ \hline       
P19         & 25           & Male       & Black         & Manager                          & USA                   & Twitter            \\ \hline       
P20         & 34           & Female     & Black         & Nurse                            & USA                   & Instagram, Twitter, TikTok           \\ \hline
P21         & 26           & Female     & Black         & Assistant                        & USA                   & Facebook, Instagram, Twitter            \\ \hline 
P22         & 25           & Male       & Middle Eastern         & Recent Graduate                  & USA                   & Facebook, Reddit, Linkedin             \\ \hline
P23         & 22           & Female     & Asian         & Student                  & USA                   & Instagram, Facebook, Reddit             \\ \hline
P24         & 21           & Male     & Asian         & Student                  & USA                   & Instagram, Twitter, Reddit             \\ \hline
P25         & 24           & Female     & White         & Unemployed                  & USA                   & Twitter, Instagram             \\ \hline
\end{tabular}
\caption{Participants' Demographic Information}
\label{table:participants}
\end{table}

\subsection{Data Collection and Analysis}
In total, we collected data through 25 semi-structured interviews with social media users. Participants were compensated with \$20 USD for their time, irrespective of the interview duration. This amount was above minimum wage at the interview time.  
Interviews lasted between 30 and 90 minutes and were all conducted remotely over Zoom. All interviews were recorded and transcribed. At the start of each interview, we asked participants for explicit verbal consent for their audio and/or video to be recorded for us to conduct analysis.
We purposely did not use any flagging interfaces as probes to avoid the focus on a single platform and instead grounded our discussions in users' prior interactions with flags on the platforms they use.
Doing so allowed our participants to speak about their experiences and concerns with flagging, which included issues with the flagging interface, but also went beyond it.

While participants were only recruited if they self-reported that they flagged content, we provided a common definition of flagging to ensure the research team and participants were on the same page. We defined flagging as ``when someone reports a post or accounts for toxic content.'' While this initial prompt was used to clarify that we were talking about the mechanism of flagging, we let our participants talk about any experiences of posts they flagged. Our emphasis in providing this context was for participants to view flagging as necessarily involving the act of clicking the report button rather than as information labels~\cite{morrow2022emerging} (which are also referred to as flags in some instances, e.g.,~\cite{figl2023symbol}) or other general aspects of encountering online harm.

During our interviews, we asked participants about the platforms they used and why they used them to understand the contexts in which they perform flagging. Next, we asked them to describe at least one instance in which they flagged content and inquired about their understanding of how flagging mechanisms operate. We prompted participants with questions aimed at stimulating wider reflection about the flagging design space, often by provoking them with hypothetical changes to flagging designs. In this process, we asked about participant concerns about privacy (e.g., the visibility of their flagging activity) to gain insights on both participant understanding of existing flagging procedueres and concerns that may result from hypothetical changes (e.g., making flags visible to friends). Finally, we explored participant experiences with the cognitive load of engaging in flagging. 
While we did not focus on a specific platform during recruiting, the three most common platforms participants discussed were Facebook, Instagram, and X.
Specific details about which platforms each participant frequented can be found in Table \ref{table:participants}.

We read and uploaded interview transcripts to Nvivo,\footnote{https://lumivero.com/} a cross-platform app for qualitative research. We applied interpretive qualitative analysis to all interview transcripts ~\cite{merriam2002}, reading and coding each interview soon after it was conducted. We ``open-coded'' ~\cite{charmaz2006} interviews in multiple iterative rounds, beginning on a line-by-line basis with codes sticking close to the transcript data.
This first round of coding generated codes such as ``wanting to understand platforms' timeline for taking action with flags'' and ``platforms should prioritize by quantity of flags.'' In subsequent rounds, we merged related codes to generate higher-level themes (e.g., a need for platform transparency about flagging processes and the necessity of flags) and identified connections between themes. Throughout coding, we also engaged in regular discussions and memo writing, which helped us achieve deeper reflections and stay alert to emerging themes. After we finished processing twenty interviews, we reassessed our data and conducted five more interviews to better flesh out themes such as ``value of flagging'' and ``responsibility for flagging.'' At that point, our analysis reached theoretical saturation, and we concluded our data collection.

\section{Findings}
In this section, we present findings on users' mental models of flagging (RQ1), motivations to flag (RQ2), and cognitive and security concerns (RQ3). We begin by exploring how users learned about flagging and their understanding of its mechanisms. Next, we discuss key motivations for flagging, such as preventing harm, social pressure, and civic duty, as well as factors that discourage flagging. We also present participants' feedback on flag design innovations, like making flags public or granting privileges to certain flag submitters. Finally, we report users' concerns about the burdens of flagging and potential security risks.

\subsection{RQ1: What are users' mental models of how social media platforms implement flagging mechanisms, and how do they develop these models?}\label{findings:understanding}
While participants shared a common understanding of the main steps involved in flagging content, they had varied interpretations of how to approach flagging due to ambiguity in platform categories and their personal preferences. 
Participants described flagging on social media platforms as involving three main steps: (1) before flagging (i.e., deciding if content should be flagged), (2) during flagging (e.g., categorizing the content and explaining why they are flagging), and (3) after flagging (i.e., receiving notifications about the outcome of their flag). 
P23 described her understanding of this process through her experiences on Facebook, Instagram, and Reddit: 
\begin{quote}
    \textit{``You stumble upon a post that you deem violates the community standards, from there platforms they have a dropdown menu...[then it] shows the different reasons why you would report such content. And then once you submit that, usually, you'd get a notification saying  `Thank you for reporting this. We'll let you know if this violates our community guidelines,' and then you wait and see.'''}
\end{quote}
While the flagging interfaces and feedback shaped participants' understanding of the process, there was variability in their approaches to each of these steps. 
In this section, we detail findings about how our participants learned about the existence of the flagging mechanism and developed an understanding of how it functions. 
We delegate the discussion of user motivations to flag to Section \ref{sec:findings-rq2}.

\subsubsection{Discovery of the option to flag content}
We found that participants first learned about the option of flagging in social settings (e.g., when with friends) or while alone (i.e., exploring site functions while on a social media platform). 
Participants sometimes began flagging after watching friends flag (P9, P11, and P17).
Others were introduced to the option in differing social settings: P8 and P15 were taught about flagging in school, and P15 was encouraged to flag racist content by his history teacher as a way of taking action: \textit{``I went to talk to her about [the racist content] and then she...explained that if someone did this to you...you could take action apart from just talking and nobody listening to you, you can actually flag.''} 
Such anecdotes show how information about flagging may be spread via real-life relationships users have. 

Several other participants indicated that they learned about flagging while exploring platform options for addressing harm. 
These participants expressed a common sentiment that on most platforms, the user interface for flagging mechanisms made it easy for them to figure out what the function was for. For example, P12 illustrated the case of Reddit:
\begin{quote}
    \textit{``The UI is so well designed that the user
    finds the report button very easily and very quickly
    as soon as they join the platform, and it's also
    good because every platform now is kind of expected to have that
    functionality. So you look for it when you want to report something.'' (P12)} 
\end{quote}

Participants also heard about the option of flagging through social media posts and news articles. 
For instance, they learned through posts directly on the platform (P2, P7, and P14) or, more specifically, from creators who mentioned flagging while discussing negative experiences with platforms moderating their content (P4 and P12). 
While these were the most common ways for our participants to hear about flagging, platform policies could also make users aware of the choice they have to flag. 
However, only P3 revealed that she personally read platform policies to learn about flagging.

\subsubsection{Review process for flagged content}
\label{sec:review_flags}
Our participants provided notably different conceptions of how they believed platforms reviewed flags. Participants' understanding differed in terms of whether they believed flagged content was automatically removed, whether there was human involvement, and if so, when it occurred and at what stage in the flagging process. While a few of our participants were confident that platforms completely automated decisions to remove content (P2 and P7), others developed varying models of how they believed platforms balanced automation and human judgment. For the rest of the participants, there was a belief that there was some group of people who reviewed flagged content on the platform side. P19 illustrates his perception of what these people do, stating: \textit{``[They] have people in place that work on the community guidelines...[they] determine if posts should be taken down or not. [They] investigate such posts.''}
While some participants shared P19's sentiment (e.g., P4 and P20), P22 believed that moderators only ``confirm'' or ``quality check'' automated decisions for whether flagged content should be removed or not due to the sheer volume of content he expected the platform to handle. 
Participants also expressed a lack of clarity regarding how platforms operationalize their guidelines to make moderation decisions.

Additionally, we surfaced other complexities in participants' understanding of how platforms review flagged content. One theory participants developed was that platforms prioritized reviewing certain categories or types of content over others. For instance, P15 stated that platforms ``prioritize some content over others...they base it off categories like severity.'' On the other hand, P17 believed platforms used an ``AI'' to determine the review priority for a post by looking at the comments and other accounts interacting with it. 

When encouraged to suggest solutions for enacting greater transparency about how platforms review flags, participants' responses alluded to a system similar to package tracking, where they would be notified immediately after they flag content that the flag submission was successful, informed about the current stage of the flag review, and then be updated about the final decision. As P3 shared: \textit{``Tell me at what point there's automated decision-making. Tell me what point there are humans. Tell me what the overall policy is. Tell me the percentage. Give me data about how many things were reported.''}

\subsubsection{Updates on flagged content from platforms}
\label{sec:outcomes_of_flagging}
After participants submitted their flags, they described a range of outcomes. In successful cases, some participants reported being notified that the platform agreed with their flags and that the flagged content was removed (P4, P6, and P7). 
However, several participants noted that platforms often failed to communicate their decisions to remove flagged content, making it unclear whether content had been removed or kept on the platform. For instance, P6, P12, and P17 assumed that the content they flagged had been removed when they noticed it no longer appeared in their feeds despite receiving no notification from the platform.
P12 reasoned that the lack of follow-up from platforms could be due to the high volume of reports, stating that \textit{``the comment just went away for me. It just disappeared and YouTube never followed up with me on it. I can't imagine they would because they must get hundreds and hundreds of reports.''}  Participants expressed frustration with where notifications were placed, as platforms could notify participants as part of their typical notification feed or in a separate portal specific to flagging. 
Regarding Facebook's flagging interface, P23 expressed: \textit{``I wish it wasn't through a separate portal...I wish you could...put out this feature where they would give you an actual notification. You can click on the notification, and it will take you to like the update.''}

However, what seemed most frustrating to our participants with the outcomes of flagging was the \textbf{inaction} they perceived on the side of platforms. 
P1 felt that \textit{``a lot of times when I flag things, they get ignored.''} What it means for a flag to \textit{get ignored}, however, varies. In some cases, participants would never receive any notifications, and no action would be taken on the flagged content. 
In other cases, platforms say they will get back to a participant but never follow up (P2 and P16). For instance, P16 noted that Twitter\footnote{We refer to the platform X as \textit{Twitter} in the Findings section because Twitter had not yet made the name change to X when we interviewed participants.}  \textit{``told [me] that they would take action and remove that post...but it's been about a week, and they haven't done anything...they should have done something a long time ago.''}
As a result of not receiving notifications from platforms and platform inaction, several participants took to monitoring content that they flagged voluntarily.

\subsubsection{Moderation actions performed alongside flagging}
After flagging content, our participants often described engaging in additional measures prior to or alongside flagging, which revealed diverse and sometimes unconventional uses of flagging mechanisms. These actions were shaped by preexisting contexts, such as participants' roles, values, and perceived platform dynamics. Some participants chose to contact the account responsible for the inappropriate content. P16 preferred to address harmful content by directly messaging the poster before resorting to flagging. He described his approach:\textit{``What I wanted to do first was inform the person nicely. Well, since I didn't get a reply...it's then time I see the effect of the flag.''} P15 used an even more proactive strategy informed by their activist identity. After flagging content on Facebook, P15 and a friend personally confronted members of an organization behind the post in person, successfully eliciting an apology and removal of the content. This suggests that preexisting social roles and personal values significantly influenced participants' responses to content-based harm. 

Other participants used existing platform interaction features to elicit greater community awareness of an instance of inappropriate content and to trigger attention for platforms to review their flag faster.
For instance, P21 recounted an instance where he commented on a Facebook video depicting the sexual assault of a young woman, calling it inappropriate. He explained his motivation: \textit{``The consequences of [me] speaking out is not as bad as those people tolerating the incidences [of abuse] that are happening.''} Additionally, P19 `liked' offensive posts because he believed they would go viral, encouraging others to flag them and pressuring the platform to review the flag faster. 

In some cases, participants developed strategies to disengage with content once they felt they had fulfilled their responsibility by flagging. For example, when the discussion became contentious in P21's example, he chose to unfollow the account, believing there was nothing more he could do. Similarly, P20 opted to block accounts, explaining that this action would \textit{“prevent them from coming up in [her] For You page.”}

\subsubsection{Misuse of flags}
Our participants were keenly aware of how flags could be misused but did not admit to misusing flags themselves. P5 noted learning about cases where users of marginalized identities who spoke out had their accounts flagged and subsequently suspended. From these anecdotes, she became worried that flagging could be an avenue for exacerbating biases against marginalized people: 
\begin{quote}
    \textit{``I've had friends and colleagues who have been targeted by hateful people who all report the poster's account for spam or something else fake, and then that person gets their account suspended...so it can be weaponized against marginalized people.''} (P5)
\end{quote}
P3 also recounted problematic past occurrences where groups of ``trolls'' on platforms such as Reddit banded together to suspend accounts they felt did not `belong' in a certain online community.
 

\subsubsection{Platform motivations to provide flagging mechanisms}
Dissatisfying experiences lead some participants to believe that flagging rarely leads to removal of flagged content and to speculate about platforms' motivations in providing users with the option to flag at all. Some participants surmised that platforms let users flag content for their own benefit. For example, P4 shared that platforms have so much content to moderate that they have little choice but to outsource the work to their users: ``\textit{There is so much data there that's being posted, that it's impossible to have like a team that moderates everything.''} Similarly, P10 speculated that platforms benefit from users flagging content: they collect flagging data and use it as feedback to improve content recommendations or platform-wide moderation by funneling information to the correct decision-maker. 
P1 and P3 suspected that legal reasons and pressure from sources such as the Congress drive platforms to offer flags.
P1 said, ``\textit{I think it's for legal and compliance reasons.''} 

Participants expressed varying degrees of trust in whether flagging as a feature is designed to truly safeguard end-users.
At one end, P20 thought that platforms offered flags as a smokescreen:\textit{``I feel like they want us to feel like we can regulate what you see. They want to make us feel like we have an option in what we see and what we interact with. We have a say in the company.''}
Along this line, P16 speculated that platforms may be motivated to ignore flags that could benefit their business models, such as not taking offensive content down because keeping it up may help \textit{``generate a lot of user traffic.''} Conversely, some participants reasoned that platforms offer flags to genuinely protect and empower users. They felt that platforms are motivated to protect and support users to keep the space they created safe (P1, P6, P12, and P18).

\subsection{RQ2: What motivates users to engage (or not engage) in flagging on social media platforms?}\label{sec:findings-rq2}
Our participants were motivated to flag content for various reasons. 
Participants' moral values guided them in deciding whether to flag specific posts. 
They further asserted that they independently made decisions to flag, even if they were pointed to potentially inappropriate content by others. 

\subsubsection{Ethical judgements in flagging decisions}\label{RQ2:ethical-judgement}
All participants flagged content on their feeds that they deemed harmful, basing their decisions on personal ethical judgments rather than strictly adhering to platform-defined guidelines. For many, flagging reflected a clear, pre-existing belief that certain content should be removed. P10 and P19, for instance, described relying on their own values to address whether the content was \textit{``right or wrong''} to flag. P19 explicitly described his conscience as the \textit{``moral standard''} for these decisions, stating 
\textit{``As long as it's up to my conscience that this is right or this is not right...I flag because, according to me, going to my judgment, this post is bad.''}

The act of flagging served primarily as a way to align participants' objections with the categories provided in the platforms' flagging interfaces rather than clarifying their own judgments. Few participants consulted community guidelines or platform policies (P3 and P15), instead opting to select categories that best matched their personal judgments. Commonly selected categories included sexual harassment (P16, P18, P21), domestic abuse (P19), impersonation (P24), spam (P12, P16, P20), nudity (P20), racism (P2, P3, P15), and sexism (P1, P25). These choices reflected users' instincts in identifying harmful posts while treating platform categories as somewhat of a framework to communicate their concerns. Despite this, participants were often uncertain about how platforms interpreted their categorizations, especially in ambiguous cases. For instance, P20 wondered how platforms distinguish between harmful content classifications, such as teasing versus outright harassment, when reviewing flags to decide what content to remove.


\subsubsection{Demands of social pressure in flagging}
Participants also admitted to flagging content when prompted by others. They described getting invited to flag via seeing others' comments, through direct messages (DMs), or hearing from friends through other channels of communication. P1 said, \textit{``So my friend had messaged us in the group thread and was like, hey, this is really bothering me. Can you all please report and join me?''}

Similarly, P6, P15, and P16 were contacted directly by people they knew who asked them to flag something. P12 noted that though she flagged comments because others urged her to, she did so only if she agreed with others' rationale. P19, however, was less strict when deciding whether to flag, claiming, ``\textit{I might flag the post without even checking it}'' when someone asks him to flag something. 
Thus, participants showed varying degrees of compliance to social pressure in their flagging behaviors.
Some participants were motivated to flag content by others who shared their identity characteristics. For example, P5 said: \textit{``I have been encouraged particularly by other natives and other queer people and trans people to go ahead and report something. Usually, it's a prominent figure who said something that really crosses a big line.''} 
Participants who were prompted to look into harmful content and flag it by others agreed to examine the content, but most emphasized that the final decision to flag the content was theirs alone. The only exceptions were cases where participants flagged to appease their friends who expressed their passion for sanctioning the post in question. In such cases, P19 and P21 said they would flag the content because they trusted their friends. P19 noted: \textit{``Because I knew she said it's bad, so I will actually flag it.''}

\subsubsection{Collective duty of flagging}
While all participants believed platforms are responsible for proactively detecting and regulating offensive content (which many erroneously referred to as platforms' ``flagging''), most also thought individual users also have a responsibility to flag content. 

More than half of our participants believed that every user is responsible for flagging norm violations. This belief was strongly tied to their reasoning to flag content to protect others. P8 expressed: \textit{``I think that it is a collective duty to keep social media space clean, and so it's everyone's responsibility to flag all inappropriate things.''} P13 and P22 shared P8's sentiments and used the analogy of cleaning to argue that everyone is responsible for contributing. P22 said:

\begin{quote}
    \textit{``If everyone does it, then the world would be a better place...If there are some plastic bottles in the street and someone picks them up and puts them in the recycling --- it is not going to remove all the plastic from the environment, save the turtles, whatever. But it's a start, and everyone should do their part no matter how small it is.''}  (P22)
\end{quote}

P9 recognized a norm of generalized reciprocity~\cite{whitham2021generalized}, arguing that flagging in \textit{``social media is not about just flagging...because what you're doing will help someone. One day you don't know if the person you're helping is your friend or your brother.''}

For some participants, this sense of responsibility was guided by their values of protecting loved ones. For example, P14 and P20 prioritized shielding their friends and family from offensive or burdensome material. As P20 explained, \textit{``It's coming to protection. I'm protecting my kids, those people that I know.''} This sense of responsibility extended to their broader community, as participants flagged content they believed could cause harm, such as racism or explicit material, with a focus on safeguarding vulnerable groups, particularly children.

\subsubsection{Flagging as an avenue for user expression within content moderation systems}
Participants valued flagging as a mechanism to offer their feedback. In fact, several expressed that flagging was a ``right'' they had as users on social media platforms (P2, P10, P19, and P20). Despite not knowing the outcome of their flags or finding that flagged content was not removed, participants remained optimistic and confident about the value of flags in expressing their voices. 

\begin{quote}
    \textit{``Maybe it's not really helpful, or maybe it's merely making me feel better. But if something is done about the fake posts, which  sometimes happened, at least in terms of Facebook,  then I'm probably helping others avoid potentially very dangerous information.''} (P3)
\end{quote}

Similarly, other participants believed in the value of flagging as a way to interact with the moderation system. P12 thought that it offered \textit{``an excellent way to moderate the space that a lot of people use every day''} and found the value in flags even greater because of the alternative, i.e., having no option to give input at all. In her words, \textit{``thinking of a world where there's no reporting, gosh...there must have been a lot of toxic language thrown around.'' }

\subsubsection{Inadequate platform responses to users' flagging efforts}\label{RQ2:compensation}
Many factors that demotivated participants from flagging revolve around their feeling that the platforms' responses to their prior flags did not adequately compensate for the data or effort they provided while flagging.
P2 received an automated message from a bot immediately after flagging content that acknowledged the flag receipt, while P15 and P20 also received some level of communication from the platforms on which they flagged content. However, none of the posts these participants flagged were taken down even weeks after they flagged them. Participants' sentiments rang strongly of frustration and feeling wronged. 
P2 and P15 used almost the exact same words about platforms communicating with empty promises. P2 said:
\begin{quote}
    \textit{``They told me that they will take action and that they will remove both, but they haven't done it up to now...It's about a week, and they haven't done anything. They should have done something a long time ago, like on the same day [it was flagged].''} 
\end{quote}

Without any feedback from platforms, participants created their own folk theories about the effects of flagging. For example, 
P4 suspected that Twitter automatically blocks accounts of flagged content, preventing users from seeing any content from that account again in their feed. Therefore, P4 felt blindsided when unable to see the results of her flagging efforts: \textit{``I don't know if it's working well because I actually don't know if it's working. I have no idea what happened to that account...I have no means to actually see the status of my reporting.''} These participants stated that the negative experiences with having their flags ignored and a lack of communication from platforms decreased their motivation and made them less likely to flag content in the future. 

Indeed, in some cases, participants' interpretations of what content is flaggable were translated as what content they were confident the platform would remove. For instance, P23 felt a post claiming that a sexual assault victim was lying should be removed because she believed that the victim in that case was telling the truth. 
However, she did not flag that post because she \textit{``felt like even pre-Elon Musk Twitter would just see that as an expression of opinion.''}
P1 developed a strategy when flagging to avoid wasting their effort:  \textit{``I have to pick and choose because I find that a lot of times when I flag, things get ignored.''}  
In a similar vein, P2, P3, and P9 developed a more grim, fatalistic take on flagging, expecting platforms to ignore most flags. 
Together, these findings make it evident that prompt and clear feedback is essential for users to feel heard and be motivated to continue flagging efforts.

\subsubsection{Strategies to flag posts versus accounts}
Participants who flagged accounts exercised caution as they viewed flagging accounts as a more serious action than flagging a post because the consequences could be more severe with potential account suspension or ban (P16, P19, and P20). 
Participants used specific strategies to determine when they should flag a post versus an account. 
Many flagged an account (as opposed to just individual posts) when they encountered extremely inappropriate singular utterances from that account and/or when they saw that account repeatedly posting inappropriate content.
For example, P19 said:
\textit{``if it's racist content, I'll definitely report the content \textbf{and} the owner.''} In contrast, P20 reported an account whenever she realized she was flagging too many inappropriate posts by that account. 
Some participants also attempted to infer the intentions of accounts posting seemingly inappropriate content. For instance, P19 noted that if he felt the intention of an account was good (e.g., promoting community awareness of sexism toward female athletes by posting examples of sexist language), he would not flag it. However, in contrast, he would immediately flag an account if he felt it had bad intentions (e.g., posting about abusing one's spouse). 

\subsection{RQ3: What are users’ cognitive and security concerns about engaging in flagging on social media platforms?}
In this section, we surface participants' concerns regarding the cognitive load and privacy concerns of flagging. Concerns about the cognitive load of flagging emerged, especially among users with marginalized identities who felt facing disproportionate exposure to harmful content. 
To interrogate this further, we investigated participant perspectives on how platforms could mitigate users' burden by prioritizing reviews of certain types of flagged content.
Finally, we describe users' privacy concerns, including apprehensions about retaliation from flagged users, which underscore the need for platform-level protections to safeguard those who participate in content moderation.

\subsubsection{Identity-based harms and cognitive load of flagging}

Participants were distinctly aware of how flagging as a mechanism places the burden of moderation on them. Those with underrepresented identities in our sample especially emphasized the additional burden of flagging they face.
For example, P1 observed that they end up seeing more harmful content due to their identity characteristics: \textit{``I will flag things because, unfortunately, if you're a queer person, if you're a trans person, if you're a person of color, if you're a woman, you will regularly encounter really heinous comments.''} 

Several participants noted that flagging takes more effort than it should. As described in Section \ref{RQ2:ethical-judgement}, categorizing harmful content can be confusing when the categories platforms provide of inappropriate content fail to cover all types of norm violations. This burden is exacerbated by platforms' differing flagging policies, resulting in different categorizations. 
As P25 proposes: 
\begin{quote}
    \textit{``I wish you could just click a button, report it, and it would just go to whatever algorithm or software that determines this does or doesn't meet our requirements [and] maybe you could have the option to comment specifically.''}
\end{quote}

The cognitive load of flagging is further exacerbated when platforms fail to act on flagged content. Participants expressed frustration that their efforts often feel futile when inappropriate content remains on the platform. P12 articulated this sentiment, stating: \textit{``Everybody has the responsibility [to flag], but the responsibility of taking action towards it is of the platforms.''} This lack of action reinforces the burden on users, who must repeatedly flag the same or similar harmful content, amplifying their sense of frustration and fatigue.

Participants also discussed potential ways platforms could alleviate this burden. P12 emphasized that platforms should at least \textit{``follow their own policy''} and be transparent in their flag review processes. Transparency could help reduce the uncertainty users feel about whether their reports are taken seriously. Similarly, P16 speculated that using automated systems or bots could reduce the strain on both users and human reviewers by handling initial reviews of flagged content. These solutions highlight the importance of reducing the effort users expend in flagging harmful content, especially for those disproportionately affected by identity-based harms.


\subsubsection{Prioritization order of reviewing flags}

Given the cognitive toll of flagging, particularly for users with marginalized identities who are disproportionately exposed to harm, prioritization strategies emerged as a key concern in our study. As noted in Section \ref{RQ2:compensation}, much participant frustration stemmed from perceived delays and inadequate responses to flagged material. With platforms tasked with reviewing vast volumes of reports, improving flag review prioritization order could result in more just distributions of benefits and burdens across flag submitters. 

To examine this, we prompted participants with the question: \textit{should users be considered equally when flagging?}
When presented with the idea of a ``super-flagger status,'' i.e., individuals given higher priority for review when they flag content, most participants said that all users should \textit{not} be considered equally, though not all agreed on the details of that prioritization. Several participants argued that platforms should prioritize reviewing the posts flagged by influential figures such as celebrities and political leaders because their popularity suggests that many users would trust their flagging actions. P15 said: 
\begin{quote}
\textit{``If they flag something right now, it should be taken serious much quicker. [The likelihood] is very high because of that large audience they reach. Basically, they have a large audience, so they should be given priority.''} 
\end{quote}

Other participants asserted that users should be prioritized based on their reporting history (P4, P5, P12, and P19). For instance, if users abuse flagging privileges, the content they flag in the future should be lower in priority for the platform to review. P19 also argued for prioritizing users' flags based on the amount of time they spend on the platform.

In contrast, some participants argued that all flag contributors should be considered equally. They offered three beliefs to support this argument: the ideal of equal rights for all suggests that all social media flaggers also be treated fundamentally equally; having more followers should not qualify one to have more privileges when flagging; and prioritizing by user as a strategy is not effective. P20 considered the situation from a business perspective:  \textit{``I feel like we are all promoting Instagram by using it. So we are all consumers, and we are all adding revenue to the company. So they should consider us equally.''}

We further explored this topic with our participants by inquiring: \textit{should types of flags be considered equally?}
Our participants' responses to this question show a lack of consensus on how flag reviews with different categories of rule violations should be prioritized.
Five participants maintained that not all categories of flagged content should be considered equally (P1, P9, P15, P16, and P17). P1 and P9 mentioned harmful content that is potentially life-threatening to be reviewed with the highest priority, while P15 thought racial discrimination should be at the top of the list. 

Many participants opposed this view and instead held that all flags should be considered equally (P2, P8, P14, P17, P18, P19, and P20). P18 had a simple yet strong reason that flags should be considered equally: \textit{``If something is not good for the community, it's not good for the community.'' }
Similarly, P8 viewed flagging itself as a simple mechanism to generate candidates for the moderator queue, saying that \textit{``flagging is just to bring attention to review it.''} 
Some participants felt that establishing a priority order for reviewing different content categories is untenable.
For example, P20 could not determine how the flagging system could prioritize handling racism vs sexism because of her firsthand experiences of being discriminated against both as a woman and a Black person.

\subsubsection{Privacy for personally identifiable information}
In addition to concerns about the cognitive burdens of flagging, participants also surfaced several privacy-related concerns. 
They expressed apprehensions about where flagged submissions go for processing and retaliation from opposing parties. Several participants did not want their personally identifiable information to be revealed when they flagged content. 
P21 expressed concern about human moderators or staff at platforms seeing their information but comforted herself with the reasoning that \textit{``these people are way too busy to be counter-checking every account.''} P5 also worried about the trustworthiness of moderators behind the scenes and speculated that those with ill intentions could get an \textit{``entry-level position''} and \textit{``pass the background check''} to become a moderator. 

In contrast, both P18 and P22 had few to no concerns about their personal information being leaked when they flagged content. P18 preferred that the platform keep information about past content he flagged so that similar content would not show up in his feed in the future.

\subsubsection{Retaliation from flagged users}
Although in the minority, some participants also related negative experiences where owners of accounts or content they flagged retaliated, causing them to feel unsafe.
 P12 developed a habit of blocking accounts after receiving hateful comments when speaking out against a post and then flagging its content: \textit{``They (account owner of flagged content) messaged me privately, and they called me names. So I ended up having to block and then report them.''}  

P12's anecdote raises an important consideration. Though flagging is anonymous at the moment, in the aftermath, an array of concerns with retaliation arise that must be examined before new features are implemented, e.g., showing one's followers the posts or accounts one flagged. 
These concerns may be exacerbated if flagging were to become more transparent. To further explore this topic, we prompted our participants with the question of: \textit{who should see flags?}

Given the hypothetical option of changing who can see when they flag content, several participants decided that all other users should be able to see when they flag content. In return, they also expected to see other users' flags. 
The prevalent sentiment for implementing this change was the belief many participants had that the more times an inappropriate post is flagged, the more likely it is to be taken down.
Per P4: \textit{``With only one flag in...it means maybe moderators will not pay attention to it, but with other people flagging it means actually some action will be taken.''} 
As such, P7 added that public flags would make platforms more accountable for promptly reviewing the flagged posts. He claimed, \textit{``If a post is being flagged, I think everybody has the right to know.''}

Other participants suggested alternative approaches, such as having only a subset of people be able to see the content they flag (P9, P3, P19, and P21). P3 preferred that only a few of her followers, such as her close friends, could see the content she flags to avoid conflict since she is followed by people she has \textit{``very different views from.''} P19 supported having his followers see the content he flags but wanted to be able to toggle this feature on and off in settings. Like P19, P21 preferred that people saw only some types of content she flags since she could be \textit{``petty''} when flagging some content but passionate at other times about flagging harmful content, depending on the topic.

\section{Discussion}

\subsection{The Need For and Importance of Flags}\label{sec:disc:importance}
Our analysis suggests that users interact with flagging mechanisms in complex ways, and ideas about how they can be improved are plentiful and diverse. While outcomes of flagging are often perceived as disappointing by users such as our participants, our analysis uncovered a strong user sentiment that flagging mechanisms are vital to their experience on social media platforms. Our participants described flagging as one of the core avenues for users to express dissatisfaction with content they deemed inappropriate. It is necessary because it provides a unique function for these users to anonymously report content directly to platforms and request to have it removed from the entire site. Besides flagging, our participants also noted using personal moderation tools, such as blocking mechanisms, to protect themselves from exposure to inappropriate content. However, our analysis highlights a distinction that users make among flagging mechanisms and other personal moderation tools like blocking: participants often flag to protect others from exposure to inappropriate posts, but they may primarily deploy blocking to protect themselves~\cite{jhaver2023personalizing}. 

Thus, flagging is not an isolated tool, but it constitutes a key component of the sociotechnical assemblages users rely on to address harm online.
Future research should examine in greater depth how users deploy it alongside or in place of other moderation mechanisms and actions. Such work may benefit from our conceptualization of the three stages of flagging as other mechanisms would present different affordances, procedural insights, cognitive burdens, and privacy considerations in their analogous temporal stages.
Further, certain moderation mechanisms are designed for the integration of specific stages of the process (e.g., blocking a user after flagging). In this way, we can also see how flagging fits into the ecosystem of moderation mechanisms, such as blocking or muting accounts.

The anonymity built into flagging mechanisms allows users to avoid retaliation by not having to confront perpetrators who post inappropriate content. In some cases where our participants attempted to avoid flagging by directly commenting on an inappropriate post or directly messaging a user (i.e., requesting the outright removal of another user's content), they were met with inappropriate comments or no response. In these instances, they resorted to flagging as it was the only other option they had to request for the content to be removed. This extends our prior understanding of how users leverage their online social connections and content moderation tools~\cite{wright2022automated,jhaver2023personalizing} to address online harms ~\cite{scheuerman2021framework,scheuerman2018safe}. 

We also found that flagging is a critical part of users' social media experiences because it is embedded in many social interactions users have on and offline. For example, our participants told stories of learning about the function of flags through word-of-mouth and seeing others flag content. These instances demonstrate that the influences and efforts that shape users' decisions to flag content stretch beyond interactions on specific social media platforms. What's more, our findings about how our participants were often encouraged to flag in an effort to protect others or to ``team up'' by mass-flagging inappropriate content further demonstrates how flagging is actively employed by users for critical social functions such as protecting their friends or reinforcing commonly shared values about what is inappropriate.
These insights add to ongoing conversations about how social media users leverage their critical postdigital literacies to create new cultural and ideological meanings of platform-implemented features like flagging ~\cite{hagh2024critical} and, more broadly, how users interact with platform governance ~\cite{chan1998procedural, ostrom1990governingthe}. In sum, flags serve as a mechanism deeply embedded into users' social media interactions, providing them a crucial opportunity to report inappropriate content anonymously. 

\subsection{Platform Versus User Responsibilities for Regulating Inappropriate Content}\label{sec:disc:responsibilities}
Our analysis suggests that users perceive flagging as a right but also, at times, an obligation. Many users viewed flagging as a \textit{right} they have that platforms should uphold because it is the only avenue available for users to express that content should be removed for everyone on the platform (see Section \ref{sec:disc:importance}). 
However, the sentiment of generalized reciprocity, along with users' inclination to take on nurturing and supportive roles in their communities~\cite{seering2022metaphors}, may also lead users to view flagging as an \textit{obligation}. 

Indeed, one of the key sentiments we surfaced through our analysis was users' personal ethical judgments motivating them to flag content, which were driven more by a moral desire to protect others from exposure to that content rather than by ensuring compliance with platform rules or policies. Users expressed motivations rooted in broader community care, such as protecting children and family members, which sometimes led them to become activists for issues such as sexual violence. While our findings highlight this sentiment in the context of general community protection, \citet{riedl2024reporting}'s recent study showcases a specific instance of users acting to safeguard a celebrity. In that case, Taylor Swift fans, motivated by a strong sense of responsibility to protect her, proactively searched for and reported non-consensual pornographic deepfakes of the pop superstar on X--a stark example of users acting in ways platforms may not anticipate or fully support.

Moreover, users' sense of responsibility to keep others safe often leads to significant cognitive burdens, as seen in prior studies on data workers who engage in moderation or annotation to protect their communities voluntarily ~\cite{dosono2019mod, matias2019civic, wohn2019volunteer, kiel2017could} and as part of their profession ~\cite{wang2022whose, zhang2024aura}. 

As we found, flagging mechanisms do not seem to be designed to consider the nuances of user motivations for flagging content. Platforms like YouTube prioritize users who flag with the most ``accuracy,''\footnote{https://blog.youtube/news-and-events/growing-our-trusted-flagger-program/} framing flagging as a mechanism for annotation rather than community care. 
This approach aligns with the metaphors of ``custodians'' or ``janitors'' applied to content moderators~\cite{gillespie2018custodians, seering2022metaphors} but overlooks the emotional and cognitive toll this labor imposes on regular end-users. This distinction underscores a mismatch between platform affordances and user expectations: while platforms rely on users to identify and annotate candidates for moderation, they fail to consistently implement flagging designs that respect users' agency and cognitive well-being. 

In contrast, our findings illustrate an image of regular users' flagging as more metaphorically aligned with that of a civilian picking up trash on the street. They may feel obligated to do so to keep their community clean and may even be encouraged by those around them, but it should ultimately be their choice to do so. We urge that improvements in flagging designs strive to align with users' model of flagging. If platforms do not improve the design of flagging systems to better align with users' motivations and mitigate their burdens, they risk placing disproportionate responsibility on users without offering adequate support. This imbalance can lead to the exploitation of users' unpaid labor, turning flagging into an extractive process rather than a collaborative mechanism for fostering community care.

\subsection{Design Implications for Improving Flags}
Through our study, we surfaced tensions and challenges within all three stages of the flagging process. We consolidate our analysis into design opportunities organized by each stage of flagging as well as key considerations and examples of feature innovations that may serve as design constraints or otherwise limitations in Table \ref{table:design-opp}. We incorporate core values we surfaced through our analysis such as the need for greater transparency and user agency in our design opportunities. These considerations are guided to provide users with satisfactory information about each stage of flagging as part of the platform's responsibilities in offering flags. Applying the knowledge that people tend to behave better when decision-making processes are perceived as fair and just ~\cite{ostrom1990governingthe}, users may be more inclined to flag if they feel that platforms will respect their time and effort in return. Without such improvements in flags, it will be difficult for platforms to continue relying on user flags to identify inappropriate content. For instance, some of our participants concluded that flagging is simply a ``smokescreen'' with no platform functionality or support, which led them to stop flagging altogether. Across the design opportunities, we provide considerations that urge platforms to build carefully designed \textit{seams}~\cite{eslami2015always,eslami2016first,jhaver2018airbnb} that inform users about how flagging works. Below, we expand on these design opportunities, highlighting the nuances that should be considered when further designing in these spaces.

\begin{table}[ht]
\begin{tabular}{|p{0.03\textwidth}|p{0.3\textwidth}|p{0.3\textwidth}|p{0.3\textwidth}|}
    \hline
    \textbf{\#} & \textbf{Design Opportunities} & \textbf{Considerations} & \textbf{Examples of Feature Innovation} \\ \hline
    \multicolumn{4}{|l|}{\textbf{Before (determining flaggability)}} \\ \hline
    1 & Flagging icons & Ensure flags are accessible (i.e., easy to see and select; not embedded within long action menus).  &  Red flag present in the upper right-hand corner of each post.\\ \hline
    2 & Increasing flagging literacy & The purpose of flagging (i.e., to report inappropriate content) should be clearly communicated. & A pinned post with clearly defined types of inappropriate content to flag. \\ \hline
    3 & Conveying information about flagged content & Further exploration of innovations to flags that decrease the anonymity of flags should be carefully explored while attending to users' diverse concerns. & Display that ``15 other users reported this post'' or personalized display of ``your friend Sally Wang reported this post.'' \\ \hline
    \multicolumn{4}{|l|}{\textbf{During (classifying flagged content)}} \\ \hline
    4 & Classification schemes for types of inappropriate content & Offer greater clarity about the meanings of included categories; ensure transparency in additions or changes to the classification scheme. & Examples of content that fits in each category.  \\ \hline
    5 & Options to provide additional information about flags & The design must balance overburdening users to include too much information (e.g., free response) and not providing enough options to indicate nuanced opinions. & Optional short-response text boxes or file upload. \\ \hline
    \multicolumn{4}{|l|}{\textbf{After (monitoring flag outcomes)}} \\ \hline
    6 & Visualization of flag review status & This system should be integrated within the platform for accessibility. & A package-tracking-like system to showcase what review stage a flag is at. \\ \hline
    7 & Visualization of platform statistics about flagging & To preserve privacy, we may urge platforms to provide aggregate statistics about the types of content that is flagged and subsequently removed or kept. & A common dashboard or monthly report sent to users. \\ \hline
\end{tabular}
 \caption{Design opportunities with corresponding considerations and examples of feature innovations, organized by the three flagging stages --- before, during, and after flagging.}
 \label{table:design-opp}
\end{table}

\subsubsection{Before Flagging}
Our analysis shows that users appreciated the ubiquity and similarity of flagging interfaces across different platforms. Their familiarity directed participants where to look for flag icons and submit reports against inappropriate content. Therefore, when designing flagging interfaces, new platforms should leverage this familiarity and deploy flagging icons, placements, and vocabulary in ways that mirror existing platforms. Our findings also suggest that platforms should try to minimize the cognitive labor involved in discovering flag icons and completing flag submissions and reduce the time required to process flags.

Some social media users may be unaware of how to use flags or not fully understand their value in addressing online harm. This lack of guidance can lead users to misuse flagging, such as by using it to simply express that they dislike content or show animosity to a user ~\cite{kou2021flag}. Since our findings show that users learn about flagging from diverse sources, such as friends and school teachers, we encourage practitioners to design ways to leverage these community interactions.  We encourage the collaboration of platforms with schools, educational leaders, media sources, and content creators to promote awareness and understanding of flags and other safety mechanisms on social media. Additionally, existing platform communications about the purpose of flagging can be improved so that they are more accessible to understand.

Crawford and Gillespie raise the question of whether ``some flags are worth more than others,'' such as flagged extremist content or content flagged by a selected group of individuals~\cite{crawford2016}. Our findings about user support for such prioritization are mixed and vary by context. For instance, users described certain cases where flags submitted by active members of a community should be prioritized, as they have a better understanding of what is inappropriate in that setting. There are existing efforts to explore this, such as Youtube's 2016 Heroes program that allows a select group of ``trusted flaggers'' to have their flagged content given priority review and have access to tools for mass-reporting content.\footnote{https://blog.youtube/news-and-events/growing-our-trusted-flagger-program/} 
This program selected users who volunteered based on how accurately they could flag content and promised to expand participation if successful. However, our findings highlight tensions between users' perceptions of what content should be removed and platforms' definitions of inappropriate content (see Section \ref{RQ2:ethical-judgement}). Using a metric for accuracy that ignores user perspectives risks alienating passionate contributors who might otherwise engage in such programs. At the same time, concerns persist about ``trolls''~\cite{phillips2015we} or malicious users abusing the flagging system to target appropriate content. Despite these issues, the program has not been publicly evaluated, nor have platforms addressed these concerns transparently.

In other cases, it was unclear what would qualify content as extreme enough to be prioritized over other types of flagged content. This presents an opportunity to further explore how the status of the person flagging and the type of flagged content should be prioritized, if at all. We recommend practitioners design such guidelines in partnership with users and carefully consider how prioritization choices would influence the addressing of user concerns across various communities and contexts of online harm. 

On the other hand, users expected that the number of people who flagged a piece of content should warrant its prioritization. Users showed interest in modifications to flag designs that would show the number of people who have previously flagged any content. However, they were concerned about the invasion of privacy when probed about further modifications that would show more specific information, such as demographics (e.g., 50\% of users that flagged this post were Asian) or personal connections (e.g., your friend Bill flagged this post). Indeed, some of our participants who engaged in flagging in a public manner (e.g., commenting that a post is inappropriate before escalating to flag) often faced retaliation in the form of threatening messages. Some platforms have already implemented flags that are more public-facing to encourage greater user involvement in identifying misinformation. For instance, on the Chinese social media platform Zihu, users flag content for it to be marked as ``controversial'' and then have the option to participate in a discussion page to raise questions about the content ~\cite{li2022flagging}. However, as 
\citet{li2022flagging} point out, only users with higher ``yan'' scores from being more active and having a greater impact on the platform as calculated by Zihu could post in these discussions.  We advise that modifications of flags are made with careful consideration of users' privacy needs, equity expectations, and audience management goals ~\cite{marwick2011tweet}.

\subsubsection{During Flagging}
In line with prior research~\cite{haimson2021disproportionate}, our participants expressed concern that individuals from marginalized groups face a greater burden of reporting inappropriate content. At the same time, these groups frequently get falsely reported. Therefore, we recommend that platforms invest in creating accurate ex-ante identification~\cite{grimmelmann2015virtues} and regulation of activities that harm these groups. Platforms could partner with these groups to better understand the nuances of online harms (e.g., the use of offensive slang words that outsiders may not recognize~\cite{jhaver2018blocklists}) and the moderation challenges they face. 

We found that users often struggle to understand the meaning and scope of inappropriate content categories within flagging interfaces. 
To address this, platforms could provide examples to illustrate the meanings of these categories, similar to what has been explored in personalized moderation mechanisms ~\cite{jhaver2023personalizing}. This approach, however, must be designed carefully to prevent bad actors from exploiting the information to evade detection. 
Additionally, we observed users struggling to express their objections in the narrow vocabulary of complaints~\cite{crawford2016} that flag interfaces provide.
Thus, platforms should enact changes that let users flag content falling outside of the existing rule violation categories more conveniently. 

Platforms such as Facebook post generalized statements about their commitment to users but have yet to elaborate on how user perspectives are actually taken into account. For example, Facebook's moderation team determines whether to remove flagged content based on its Community Standards\footnote{https://transparency.meta.com/policies/community-standards/}, which it claims are ``based on feedback from people and advice from experts.'' However, it remains unclear how this feedback is gathered--particularly whether users with marginalized identities, who are more likely to experience harm, receive any specific support.
Additionally, our findings suggest that users lack a concrete understanding of how such community standards are operationalized to make content moderation decisions. 
Offering insights into such procedures could help platforms build trust with flag submitters.

No aspect of flagging should be overlooked when considering improving the design of flags to reduce the burden of labor placed on users. As our findings confirm, users view the act of flagging as a type of labor and may be less inclined to flag content if they view the task as too labor-intensive.  For instance, X\footnote{We are referring to the current flagging interface on X in December 2024} requires users to click through multiple options when flagging, asking for free-response explanations about ``additional information'' about the issue being reported as well as for users to flag additional posts from the account they are reporting. X claims this process helps ``provide better context to evaluate your report.''\footnote{https://help.x.com/en/rules-and-policies/x-report-violation} However, this design, which maximizes user justification without careful consideration of its burden, risks discouraging users from flagging altogether. Thus, we advise that labor-intensive options that require long-form answers or the uploading of information, e.g., screenshots, are optional and implemented with a clear purpose, e.g., encouraging users to provide additional context for flagged content outside the existing categories of inappropriate content. 

\subsubsection{After Flagging}
Transparency in flagging procedures becomes increasingly critical as platforms rely more heavily on automation or related technologies (e.g., LLMs) to manage the overwhelming volume of flagged content. While users are open to the use of automation, particularly when their flags receive little attention or follow-up, they strongly prefer clarity about how automated systems and human reviewers collaborate. Our findings reveal curiosity about the division of labor within the flagging pipeline, such as who determines if the content is flagged in the correct category, when humans oversee automated decisions, and how sanctions are applied. Greater transparency into these processes can help users better understand platform constraints, manage their expectations, and potentially reduce flag misuse. Additionally, given the greater trust users tend to place in human review, platforms should approach automation cautiously, ensuring that human oversight remains a central component of their flag review systems.

This emphasis on transparency extends to the broader experience of flagging, as users also want to feel that their efforts are respected. At the moment, users' lack of ability to track the flagging status on most platforms meant that even after flagging a post, they have to check the removal status of posts they reported repeatedly.  Some platforms promise to notify users when an action will be taken on their flags but fail to do so. Users can become disillusioned with the flagging process and doubt that platforms would take any action in these cases. 

We recommend that platforms institute information and visualization systems that allow users to easily monitor the status of their flags.  
This need for improved systems aligns with prior research emphasizing the importance of transparency in platform processes. For instance, the lack of transparency has been noted as a critical concern in content removal decisions ~\cite{jhaver2019survey, jhaver2019explanations, brunk2019effect} and broader platform moderation practices ~\cite{juneja2020through}. However, most platforms either lack post-flagging dashboards entirely or rely on rudimentary interfaces, with Facebook being a notable exception—though even Facebook's system is limited in transparency and hosted separately from the main platform. We also note that YouTube's Heroes program mentioned earlier also provided users with tools to ``track their own contributions and see their overall impact.''\footnote{We were unable to find any documentation of what the dashboard looks like aside from a screenshot of flagging forum questions.} However, access to this dashboard is limited to those who volunteer and are selected for the program. While there are complexities in other tools provided in the Heroes program (e.g., mass-reporting tool), it is unclear why YouTube or other platforms could not simply provide a dashboard to track flag status to a broader selection of users, as this program was launched several years prior to our study. Overall, we find that these examples of inconsistent implementation with a lack of evaluation of such dashboards present significant opportunities for innovation in the design of post-flagging mechanisms.

In addition to technological and design improvements, the push for transparency ties into broader calls for policy reform. Scholars have highlighted the need for regulating content moderation practices ~\cite{langvardt2017regulating, buckley2022censorship} and explored how existing policies, such as the EU Digital Services Act, address problematic platform behaviors like shadowbanning ~\cite{diaz2021double, myers2018censored} and unjust moderation standards ~\cite{diaz2021double}. Such legal and regulatory attention should extend to flagging reforms, ensuring platforms are held accountable for transparent and equitable content moderation.

Finally, privacy concerns remain central to discussions about increasing transparency in flagging mechanisms. To balance these concerns with the need for transparency, we propose the design opportunity of providing visualizations for aggregated statistics about flagged content, which offers users meaningful insights while safeguarding individual privacy.

\subsection{Limitations and Future Work}

In this study, we conducted semi-structured interviews with 25 social media users who recently engaged in flagging. Though our data helped us examine nuanced user experiences and tensions with flagging in-depth, the qualitative nature of our study design and the size of our sample limited us from generalizing our results and learning about the relative popularity of various design choices in flagging interfaces. 
We also acknowledge potential desirability bias in our interviews, as participants may have avoided discussing the misuse of flags to present themselves more favorably ~\cite{bergen2020everything}. Moreover, our findings may be limited because online behaviors and perceptions may change over time. Future research could investigate factors not addressed in our study, such as the influence of education and digital literacy on mental models of flagging. Quantitative and longitudinal methods may further explore user perceptions of flag prioritization, control mechanisms, cognitive labor, and privacy protections, offering broader insights into effective flagging system design. 

We focused on the perspectives of end-users who have engaged in flagging in this work in our effort to provide empirical data that contributes to Crawford and Gillespie's existing conceptualization of flagging ~\cite{crawford2016}. It would be valuable to examine the dynamics and challenges of flag processing from the perspective of platform administrators and moderators. For example, it might be insightful to learn how false flagging is identified and sanctioned and design solutions that discourage users from submitting flags in bad faith.

We also concur with and acknowledge critical work highlighting the potential misuse of content moderation mechanisms and flagging in particular. Those with marginalized identities are most likely to fall victim to such abuses of flagging ~\cite{musgrave2022experiences}. The misuse may occur by platforms unjustly censoring or hiding flagged content~\cite{myers2018censored} or result from actions by individual users. It was for this purpose that we intentionally recruited participants with perspectives as members of marginalized communities. The design considerations we provide highlight important concerns when redesigning flags in this context. For instance, while it may be productive to experiment with non-anonymous flags, we must consider what harm public views of flagging information can result in.

Our analysis shows that users develop decision-making guidelines based on their values within the limited vocabulary of flags. Flagging can be socially motivated for some, who consider or even flag content without deeply examining it because they value the opinion of another person who asked them to flag. This distance between platforms' and users' notion of flaggability aligns with prior research on flagging motivations in online gaming communities~\cite{kou2021flag}. Examining this distance systematically could be an exciting avenue for sharpening platforms' moderation policies and practices to better align with users' values. 

Finally, the labor users provide to platforms in the form of flagging shares similarities with the labor of data labeling ~\cite{gray2019ghost, morrow2022emerging} in its voluntary nature and the potential for exposure to psychologically harmful content. Particularly as applications of large language models become more prevalent, we envision possible applications of flagging mechanisms to allow users to report harmful model output (e.g., adding the flag option to ChatGPT). Our findings could inform the design of such applications. 

\section{Conclusion}
At a time in which user participation in social media moderation is being recognized as increasingly vital~\cite{jhaver2023personalizing,jhaver2022filterbuddy,vaccaro2021contestability}, our study serves as a starting point for exploring how users report inappropriate content and how mechanisms like flags can be designed to better facilitate these interactions. Our research examined why end-users flag, their experiences with flagging mechanisms, their mental models of the flagging pipeline, and reasons why they may hesitate to flag.  We found that users perceive flags not as a stop-gap solution that compensates for platforms' limited moderation resources but as an essential avenue to voice their disagreements with the curated content. Our participants' willingness to contribute their time, even going as far as personally contacting the people who post content they flag to take it down, shows users' investment in addressing online harms. However, we found a lack of procedural transparency in flagging implementations undermines users' trust in them. We call for further exploration of how flagging interfaces can be better designed and integrated with other moderation tools to create more transparent and effective ecosystems for addressing online harm.



\bibliographystyle{ACM-Reference-Format}
\bibliography{moderation_references,refs,Flagging}



\end{document}